\documentclass[10pt,conference]{IEEEtran}
\usepackage{cite}
\usepackage{graphicx}
\usepackage{amsmath}
\usepackage{bm}
\usepackage{url}
\usepackage{setspace}
\usepackage{multicol}

\renewcommand{\baselinestretch}{.95}

\begin{document}
\title{Performance Analysis of Cell-Free Massive MIMO Systems with Massive Connectivity}
\author{\IEEEauthorblockN{Mangqing Guo and M. Cenk Gursoy}
\IEEEauthorblockA{Department of Electrical Engineering and Computer Science, Syracuse University, Syracuse, NY 13244.
\\
Email: mguo06@syr.edu, mcgursoy@syr.edu
}
}
\maketitle

\begin{abstract}
In this paper, we investigate the performance of cell-free massive MIMO systems with massive connectivity. With the generalized approximate message passing (GAMP) algorithm, we obtain the minimum mean-squared error (MMSE) estimate of the effective channel coefficients from all users to all access points (APs) in order to perform joint user activity detection and channel estimation. Subsequently, using the decoupling properties of MMSE estimation for large linear systems and state evolution equations of the GAMP algorithm, we obtain the variances of both the estimated channel coefficients and the corresponding channel estimation error. Finally, we study the achievable uplink rates with zero-forcing (ZF) detector at the central processing unit (CPU) of the cell-free massive MIMO system. With numerical results, we analyze the impact of  the number of pilots used for joint activity detection and channel estimation, the number of APs, and signal-to-noise ratio (SNR) on the achievable rates.
\end{abstract}

\begin{spacing}{1.1}

\section{Introduction}
Massive connectivity and sporadic activity are two key characteristics of future Internet-of-Things (IoT) networks \cite{LiuYu2018}. Large amount of devices (such as vehicles, sensors, smart furniture, etc.) are potential users of the IoT network, while only a small part of the users actively connect to the network at a given time instant. Active users initially send pilots and then connect to the network to exchange information within the network, and then disconnect from the network after information exchange is completed. In this way, throughput of the network can be significantly improved, and many more users could be served by the network. However, there is a fundamental challenge in networks with massive connectivity: how to recognize users which need to communicate within the network among the entire set of users? This is also known as the sparse activity detection problem.

Approximate message passing (AMP) is an efficient algorithm for statistical estimation in high-dimensional problems such as compressed sensing \cite{DonohoMalekiMontanari2009,DonohoMalekiMontanari2010,DonohoMalekiMontanari2010a}. Indeed, AMP has been widely studied and used in solving compressed sensing problems in linear systems \cite{KimChangJungEtAl2011,ZinielSchniter2013,BayatiMontanari2011}. Additionally, several recent studies have focused on solving joint activity detection and channel estimation problems in networks with massive connectivity using the AMP algorithm \cite{ChenSohrabiYu2018,LiuYu2018,ChenSohrabiYu2018a,GuoGursoy2020,GuoGursoy2019}. Compared with other compressed sensing algorithms (such as iterative thresholding algorithm \cite{BlumensathDavies2009}, compressive sampling matching pursuit algorithm \cite{HuangTaoXiangEtAl2017}, gradient pursuit algorithm \cite{ZhouZhaoHu2010}), the advantage of AMP algorithm lies in its low computational complexity and the fact that it allows efficient performance analysis. With the AMP algorithm, the received signal is decomposed into Gaussian noise corrupted scalar versions, and the performance analysis is relatively easy to perform with the state evolution equations. However, some disadvantages also exist in the AMP algorithm. For instance, AMP algorithm can diverge if the measurement matrix has non-zero mean components, or the condition number of the measurement matrix is large\cite{CaltagironeZdeborovKrzakala2014}. In order to overcome such divergence in the AMP algorithm, generalized approximate message passing (GAMP) algorithm is proposed in \cite{Rangan2011}, and damping and mean removal techniques are employed in \cite{VilaSchniterRanganEtAl2015,RanganSchniterFletcherEtAl2014} to further promote the convergence of the GAMP algorithm.

Cell-free massive MIMO is another key technology for next generation wireless systems, and it has also been intensively studied in recent years. In these systems, a large number of access points (APs) are spatially distributed in the coverage area, and all the users are jointly and simultaneously served by all of the APs. Spectral and energy efficiency analysis of cell-free massive MIMO systems with zero-forcing (ZF) detector have been conducted in \cite{ZhangWeiBjrnsonEtAl2017} and \cite{LiuLuoChenEtAl2020}. Power control in cell-free massive MIMO systems has been studied in \cite{NgoAshikhminYangEtAl2017} and \cite{BuzziZappone2017}.

Moreover, we recently studied sparse activity detection in cell-free massive MIMO systems with massive connectivity via AMP algorithm and likelihood ratio test in \cite{GuoGursoy2020}, where it is shown that the error probability in activity detection tends to zero when the number of APs, pilots and users tend to infinity while the ratio of the number of pilots and users is kept constant. Since GAMP is an extended version of the AMP algorithm, it could also be combined with the likelihood ratio test for activity detection in cell-free massive MIMO systems with massive connectivity. Furthermore, we can also conclude that the error probability in activity detection with GAMP and likelihood ratio test tends to zero when the number of APs, pilots and users tend to infinity while the ratio of the number of pilots to the number of users is kept constant. 

Our analysis in this paper is divided into two parts. First, we analyze the joint activity detection and channel estimation using the pilots transmitted from active users via GAMP algorithm and likelihood ratio test. Minimum mean-squared error (MMSE) estimates of the effective channel coefficients from all users to all APs can be obtained with the GAMP algorithm. With the decoupling properties of MMSE estimation for large linear systems\cite{DongningGuo2005}, the received pilot signals are decomposed into scalar additive complex Gaussian noise corrupted versions. We determine the variance of the scalar additive complex Gaussian noise (which is also subsequently referred to as the converged noise variance) with the help of the state evolution equations of the GAMP algorithm. Then, we identify the variances of both the estimated channel and the corresponding channel estimation error. Secondly, we address the uplink data transmission from active users. With the results in \cite{LiuLuoChenEtAl2020}, we analyze the achievable uplink data transmission rates in cell-free massive MIMO systems with massive connectivity when zero-forcing (ZF) detector is employed. In this setting, we demonstrate the impact of the accuracy of activity detection on the achievable rates via numerical results.

\section{System Model}
We consider a cell-free massive MIMO system that consists of $M$ APs and $N$ single antenna users. All the APs and users are uniformly distributed in a circular area with radius $R$, and all the APs are connected to the central processing unit (CPU) through a backhaul network. Let us denote by $d_{\textit{ij}}$ the distance from the $i$th user to the $j$th AP. The probability density function (PDF) of $d_{\textit{ij}}$ is
\begin{equation}\label{equ1}
p({d_{\textit{ij}}}) = \frac{{4{d_{\textit{ij}}}}}{{\pi {R^2}}}\left[ {{{\cos }^{ - 1}}\left( {\frac{{{d_{\textit{ij}}}}}{{2R}}} \right) - \frac{{{d_{\textit{ij}}}}}{{2R}}{{\left( {1 - \frac{{d_{\textit{ij}}^2}}{{4{R^2}}}} \right)}^{\frac{1}{2}}}} \right]
\end{equation}
for $0 < {d_{\textit{ij}}} < 2R$ and $p({d_{\textit{ij}}})=0$ otherwise \cite{Mathai1999}.

We consider flat fading channel between all the users and APs in this paper. Besides, massive connectivity and sparse activity are assumed, i.e., there is a large number of users uniformly distributed in the area, while the number of active users at a given time instant is small. We call the probability of each user being active at a given time instant as the activity probability, and assume that the activity probability of all the users is the same and denoted by $\lambda$. Moreover, the activities of all the users keep unchanged during each channel coherence interval.
At a given time instant, we denote the activity of the $i$th user by the binary-valued $a_i$. Specifically, $a_i=1$ stands for the $i$th user being active, while $a_i=0$ denotes inactivity. Then, the activity probability is $p({a_i} = 1) = 1 - p({a_i} = 0) = \lambda $.

The channel coefficient from the $i$th user to the $j$th AP is
\begin{equation}\label{equ2}
{g_{\textit{ij}}} = \beta _{\textit{ij}}^{1/2}{h_{\textit{ij}}}
\end{equation}
where ${\beta _{\textit{ij}}}$ is the large-scale fading coefficient which can be expressed as ${\beta _{\textit{ij}}} = \min \left( {d_{\textit{ij}}^{ - \alpha },d_0^{ - \alpha }} \right)$, ${d_0}$ is the reference distance, and $\alpha $ is the path loss decay exponent \cite{ChenBjrnson2018}. Since $\beta _{\textit{ij}}$ changes slowly over time, it can be measured in advance and we assume that it is known at the CPU and all the APs. With the PDF of $d_{\textit{ij}}$ as given in (\ref{equ1}), we can obtain the PDF of $\beta _{\textit{ij}}$ and denote it by $p(\beta )$. ${h_{\textit{ij}}} \sim {\cal C}{\cal N}(0,1)$ is the small scale fading coefficient.

We assume that there are $T$ symbols in each channel coherence interval. During each channel coherence interval, $L$ symbols are used for joint activity detection and channel estimation, and the remaining $T-L$ symbols are used for uplink data transmission from the users to APs. We call the former as the joint activity detection and channel estimation phase, and the latter as the data transmission phase. Furthermore, we assume the total bandwidth of the system is $B$ Hz.

We denote the combination of the $i$th user's activity and the channel coefficient from the $i$th user to the $j$th AP as the effective channel coefficient ${\theta _{ij}}$, i.e., ${\theta _{ij}} = {a_i}{g_{ij}}$. In order to simplify the description, we represent the effective channel coefficients from all the users to the $j$th AP by the $N \times 1$ dimensional vector ${{\bm{\theta }}_j}$, whose $i$th element is ${\theta _{ij}}$. Then, in the joint activity detection and channel estimation phase, the received pilot signal at the $j$th AP can be expressed as
\begin{equation}\label{equ3}
{{\bf{y}}_{p,j}} = {\bf{\Phi }}{{\bm{\theta }}_j} + {{\bf{n}}_{p,j}}
\end{equation}
where ${\bf{\Phi }}$ is the pilot matrix, and ${{\bf{n}}_{p,j}} \in {\cal C}{\cal N}(0,{\sigma _0^2}{{\bf{I}}_L})$ is the independent and identically distributed additive white Gaussian noise at the $j$th AP. We assume that the elements of the $L \times N$ complex pilot matrix ${\bf{\Phi }}$ are independent and circularly symmetrically distributed with zero mean and variance $1/L$. We further assume that both the number of pilots and number of users grow without bound with their ratio kept constant, i.e., $\mathop {\lim }\limits_{N \to \infty } \frac{N}{L} = \gamma $.

Let us also denote the activity support of all the users, i.e., the set of the indices of active users, by ${\cal S}$. Moreover, we represent the channel coefficients from the $i$th user to all the APs by the $M \times 1$ dimensional vector ${{{\bm{\varphi }}_i}}$, whose $j$th element is $g_{\textit{ij}}$. Now, in the uplink data transmission phase, the received signals at all the APs form an $M \times 1$ vector and can be expressed as
\begin{equation}\label{equ4}
{{\bf{y}}_d} = \sum\limits_{i \in {\cal S}} {{{\bm{\varphi }}_i}{x_i}}  + {{\bf{n}}_d}
\end{equation}
where $x_i$ is the data transmitted from the $i$th user, and ${\bf{n}}_d\in {\cal C}{\cal N}(0,{\sigma _0^2}{{\bf{I}}_M})$ is the noise vector whose components represent the independent and identically distributed Gaussian noise terms at the APs. We assume that the transmitted signals from all the users satisfy $E\left\{ {{{\left| {{x_i}} \right|}^2}} \right\} = 1$.

\section{Performance Analysis}
The GAMP algorithm proposed in \cite{Rangan2011} is used for joint activity detection and channel estimation in this paper. Specifically, with GAMP algorithm, we obtain the MMSE estimates of the channel coefficients. We note that GAMP algorithm decomposes the received pilot signal into scalar Gaussian noise corrupted versions, and the noise variance can be obtained with the state evolution equations of the GAMP algorithm. Following this approach, we identify the statistical characterizations of the estimated channel coefficients and the corresponding channel estimation error.

In the data transmission phase, the estimated channel coefficients are used for ZF detection. With the results in \cite{LiuLuoChenEtAl2020}, we analyze the achievable rate of cell-free massive MIMO systems with massive connectivity.
\subsection{Joint Activity Detection and Channel Estimation}
In this section, we address the decoupling principle of the GAMP algorithm, and analyze the joint activity detection and channel estimation phase.

Using the replica symmetric postulated MMSE decoupling properties given in \cite{DongningGuo2005} and \cite{Rangan2012}, we can obtain the following decoupling principle of the GAMP algorithm for cell-free massive MIMO systems:
\begin{equation}\label{equ5}
{\widehat y_{p,\textit{ij}}} = {\theta _{\textit{ij}}} + {\sigma _{\textit{eff}}}n
\end{equation}
where ${\widehat y_{p,\textit{ij}}}$ is the GAMP decoupled pilot signal, and $n \sim {\cal C}{\cal N}(0,1)$ is the additive white Gaussian noise.

Since ${\theta _{\textit{ij}}} = {a_i}{g_{ij}}$, the PDF of ${\theta _{\textit{ij}}}$ is
\begin{equation}\label{equ6}
p({\theta _{\textit{ij}}}) = (1 - \lambda )\delta ({\theta _{\textit{ij}}}) + \lambda {\cal C}{\cal N}({\theta _{\textit{ij}}};0,{\beta _{\textit{ij}}})
\end{equation}
where $\delta ( \cdot )$ is the Dirac delta function, and ${\cal C}{\cal N}(\cdot;a,b)$ denotes the PDF of a circularly symmetric complex Gaussian random variable with mean $a$ and variance $b$. Then, the MMSE estimate of ${\theta _{\textit{ij}}}$ is
\begin{equation}\label{equ7}
{\widehat \theta _{\textit{ij}}} = G\left( {{{| {{{\widehat y}_{p,\textit{ij}}}} |}^2};\sigma _{\textit{eff}}^2,\lambda ,{\beta _{\textit{ij}}}} \right) \frac{{{\beta _{\textit{ij}}}{{\widehat y}_{p,\textit{ij}}}}}{{{\beta _{\textit{ij}}} + \sigma _{\textit{eff}}^2}}
\end{equation}
where
\begin{equation}\label{equ8}
G\left(  {{{| {{{\widehat y}_{p,\textit{ij}}}} |}^2};\sigma _{\textit{eff}}^2,\lambda ,{\beta _{\textit{ij}}}} \right)  = \frac{1}{{1 + \frac{{(1 - \lambda )({\beta _{\textit{ij}}} + \sigma _{\textit{eff}}^2)}}{{\lambda \sigma _{\textit{eff}}^2}}\exp \left( { - \frac{{{\beta _{\textit{ij}}}{{\left| {{{\widehat y}_{p,\textit{ij}}}} \right|}^2}}}{{\sigma _{\textit{eff}}^2({\beta _{\textit{ij}}} + \sigma _{\textit{eff}}^2)}}} \right)}}.
\end{equation}

The state evolution equation of the GAMP algorithm is
\begin{equation}\label{equ9}
\widehat r_{\textit{ij}}^t = {\theta _{\textit{ij}}} + \sqrt {{\xi ^t}} n
\end{equation}
where $\widehat r_{\textit{ij}}^t$ is the decoupled received pilot signal at the $t$th iteration of the GAMP algorithm, and the noise variance $\xi ^t$ satisfies the following state evolution equation \cite{Rangan2011}:
\begin{equation}\label{equ10}
{\xi ^{t + 1}} = \sigma _0^2 + \gamma E\left\{ \left| {{\theta _{\textit{ij}}} - \widehat \theta _{\textit{ij}}^t} \right|^2 \right\}
\end{equation}
where the expectation (denoted by $E\{  \cdot \} $) is over both large-scale and small-scale channel fading, and $\widehat \theta _{\textit{ij}}^t$ is the MMSE estimate of $\theta _{\textit{ij}}$ based on $\widehat r_{\textit{ij}}^t$ in (\ref{equ9}), which can be obtained using the formulation in (\ref{equ7}). When the GAMP algorithm converges, we have ${\xi ^{t+1}}={\xi ^t} = \sigma _{\textit{eff}}^2$. Thus, we can obtain
\begin{equation}\label{equ11}
E\left\{ {\left| {{\theta _{\textit{ij}}} - \widehat \theta _{\textit{ij}}^t} \right|^2} \right\} = \lambda\left(  {E\{ \beta \}  -\int\limits_{{\beta _{\min }}}^{{\beta _{\max }}} {\int\limits_0^\infty  {F({\sigma _{\textit{eff}}},\beta ,t)dt} d\beta } }\right)
\end{equation}
where
\begin{equation}\label{equ12}
F({\sigma _{\textit{eff}}},\beta ,t) = \frac{{\lambda t p(\beta) {e^{ - t\sigma _{\textit{eff}}^2/\beta }}}}{{{{\left( {\frac{\beta }{{\sigma _{\textit{eff}}^2}} + 1} \right)}^2}\left( {\frac{\lambda }{{\beta  + \sigma _{\textit{eff}}^2}} + \frac{{(1 - \lambda ){e^{ - t}}}}{{\sigma _{\textit{eff}}^2}}} \right)}},
\end{equation}
${\beta _{\min }} = {(2R)^{ - \alpha }}$, and ${\beta _{\max }} = d_0^{ - \alpha }$. Then, we have the following fixed point equation when the GAMP algorithm converges:
\begin{equation}\label{equ13}
\sigma _{\textit{eff}}^2 = \sigma _0^2 + \gamma E\left\{ {\left| {{\theta _{\textit{ij}}} - \widehat \theta _{\textit{ij}}^t} \right|^2} \right\}.
\end{equation}
By solving the fixed point equation in (\ref{equ13}), we can find the noise variance $\sigma _{\textit{eff}}^2$ of the GAMP decoupled pilot signal. With this, we can determine the MMSE estimate of the effective channel coefficients, $\theta _{\textit{ij}}$, using (\ref{equ7}). Now for active users, it is immediate that
\begin{align}\label{equ14}
\textit{var}\left( {{{\widehat \theta }_{\textit{ij}}}} \right) &= {E_1}\left\{ {\left( {{{\widehat \theta }_{\textit{ij}}} - {E_1}\left\{ {{{\widehat \theta }_{\textit{ij}}}} \right\}} \right){{\left( {{{\widehat \theta }_{\textit{ij}}} - {E_1}\left\{ {{{\widehat \theta }_{\textit{ij}}}} \right\}} \right)}^*}} \right\}  \nonumber\\
&  = \frac{{\beta _{\textit{ij}}^2}}{{{\beta _{\textit{ij}}} + \sigma _{\textit{eff}}^2}}
\end{align}
where $\textit{var}(z)$ is the variance of $z$, and ${E_1}\{  \cdot \} $ stands for the expectation over small-scale channel fading. We define the channel estimation error as
\begin{equation}\label{equ15}
{\varepsilon _{\textit{ij}}} = {\widehat \theta _{\textit{ij}}} - {\theta _{\textit{ij}}}.
\end{equation}
Then,
\begin{align}\label{equ16}
\textit{cov}\left( {{\varepsilon _{\textit{ij}}},{{\widehat \theta }_{\textit{ij}}}} \right) &= {E_1}\left\{ {\left( {{\varepsilon _{\textit{ij}}} - {E_1}\left\{ {{\varepsilon _{\textit{ij}}}} \right\}} \right){{\left( {{{\widehat \theta }_{\textit{ij}}} - {E_1}\left\{ {{{\widehat \theta }_{\textit{ij}}}} \right\}} \right)}^*}} \right\}  \nonumber\\
&  = 0
\end{align}
where $\textit{cov}\left( {{z_1},{z_2}} \right)$ denotes the covariance of $z_1$ and $z_2$. Therefore, $\varepsilon _{\textit{ij}}$ and ${\widehat \theta }_{\textit{ij}}$ are uncorrelated. Recall that we assume the error probability of activity detection is zero in this paper, which equals to the case that the user activities are known as priori. Since only active users are considered during the data transmission phase ($\lambda$ can be regarded as 1), $\varepsilon _{\textit{ij}}$ and ${\widehat \theta }_{\textit{ij}}$ are complex Gaussian random variables based on the results in (\ref{equ7}) and (\ref{equ15}). Thus, $\varepsilon _{\textit{ij}}$ and ${\widehat \theta }_{\textit{ij}}$ are independent of each other. Then, we can obtain
\begin{equation}\label{equ17}
\textit{var}({\varepsilon _{\textit{ij}}}) = \frac{{{\beta _{\textit{ij}}}\sigma _{\textit{eff}}^2}}{{{\beta _{\textit{ij}}} + \sigma _{\textit{eff}}^2}}.
\end{equation}

As noted in the introduction, the error probability in activity detection with GAMP algorithm and likelihood ratio test tends to zero as $M$ tends to infinity and $\mathop {\lim }\limits_{N \to \infty } \frac{N}{L} = \gamma $, which is assumed to be satisfied in this paper. Thus, we will not consider the effect of activity detection errors on the performance analysis in terms of achievable rates in uplink data transmission in the next subsection, while the influence is studied via numerical simulations subsequently. Due to space limitations, we refer to \cite{GuoGursoy2020} for more details on the activity detection with the GAMP algorithm and likelihood ratio test.

\subsection{Achievable Rates of Uplink Data Transmission}
For active users, we have ${\theta _{\textit{ij}}} = {g_{\textit{ij}}}$. Therefore, ${g_{\textit{ij}}}={\theta _{\textit{\textit{ij}}}} = {\widehat \theta _{\textit{\textit{ij}}}} - {\varepsilon _{\textit{\textit{ij}}}}$. Substituting ${g_{\textit{ij}}}$ into (\ref{equ4}), we can obtain
\begin{equation}\label{equ18}
{{\bf{y}}_d} = \sum\limits_{i \in {\cal S}} {{{\widehat {\bm{\theta }}}_i}{x_i}}  - \sum\limits_{i \in {\cal S}} {{{\bm{\varepsilon }}_i}{x_i}}  + {{\bf{n}}_d}
\end{equation}
where ${\widehat {\bm{\theta }}_i} = [{\widehat \theta _{\textit{i1}}},{\widehat \theta _{\textit{i2}}}, \cdots ,{\widehat \theta _{\textit{iM}}}]^T$ and ${{\bm{\varepsilon }}_i} = [{\varepsilon _{\textit{i1}}},{\varepsilon _{\textit{i2}}}, \cdots ,{\varepsilon _{\textit{iM}}}]^T$.

Let us denote the estimated effective channel coefficients from all active users to the entire AP set by the $\left| {\cal S} \right| \times M$ matrix ${\bm{{\rm B}}}$, i.e., the element on the $i$th row and $j$th column of ${\bm{{\rm B}}}$, ${b_{\textit{ij}}}$, equals ${\widehat \theta _{\textit{ij}}}$ for ${i \in {\cal S}}$. For uplink data transmission, we employ the ZF receiver at the CPU, i.e., the detector matrix is ${\bm{\Omega }} = {\bm{{\rm B}}}{\left( {{{\bm{{\rm B}}}^H}{\bm{{\rm B}}}} \right)^{ - 1}}$. Then, the detected signal is
\begin{align}\label{equ19}
{\widehat x_k} &= {\bm{\omega }}_k^H{{\bf{y}}_d}  \nonumber\\
&  = \sum\limits_{i \in {\cal S}} {{\bm{\omega }}_k^H{{\widehat {\bm{\theta }}}_i}{x_i}}  - \sum\limits_{i \in {\cal S}} {{\bm{\omega }}_k^H{{\bm{\varepsilon }}_i}{x_i}}  + {\bm{\omega }}_k^H{{\bf{n}}_d}
\end{align}
where ${{\bm{\omega }}_k}$ is the $k$th column of ${\bm{\Omega }}$. Now, using the fact that Gaussian noise is the worst-case noise \cite{HassibiHochwald2003}, the achievable uplink rate can be obtained as
\begin{equation}\label{equ20}
R = \frac{{T - L}}{T}B\widetilde R
\end{equation}
where
\begin{equation}\label{equ21}
\widetilde R = \sum\limits_{i \in {\cal S}} {{E_1}\left\{ {{{\log }_2}\left( {1 + \frac{1}{{\sum\limits_{j \in {\cal S}} {{{\left| {{\omega _{\textit{ij}}}} \right|}^2}\sum\limits_{k \in {\cal S}} {\textit{var}\left( {{\varepsilon _{\textit{jk}}}} \right)} }  + \left\| {{{\bf{\omega }}_k}} \right\|_2^2}}} \right)} \right\}}.
\end{equation}
Next, we can determine a lower bound $\widehat R$ for the achievable uplink rate as follows \cite{LiuLuoChenEtAl2020}:
\begin{equation}\label{equ22}
\widehat R = \frac{{T - L}}{T}B\sum\limits_{i \in {\cal S}} {{{\log }_2}\left( {1 + \frac{1}{{{\varepsilon _{\max }} + 1}}{\varsigma _i}\left( {{\xi _i} - 1} \right)} \right)}
\end{equation}
where
\begin{equation}\label{equ23}
{\varepsilon _{\max }} = \max \sum\limits_{i \in {\cal S}} {\textit{var}\left( {{\varepsilon _{\textit{ij}}}} \right)},
\end{equation}
\begin{equation}\label{equ24}
{\varsigma _i} = \frac{{\sum\limits_{m \in {{\cal M}_k}} {{{\left( {\textit{var}\left( {{{\widehat \theta }_{\textit{im}}}} \right)} \right)}^2}} }}{{\sum\limits_{m \in {{\cal M}_k}} {\textit{var}\left( {{{\widehat \theta }_{\textit{im}}}} \right)} }},
\end{equation}
\begin{equation}\label{equ25}
{\xi _i} = \frac{{{{\left( {\sum\limits_{m \in {{\cal M}_k}} {\textit{var}\left( {{{\widehat \theta }_{\textit{im}}}} \right)} } \right)}^2}}}{{\sum\limits_{m \in {{\cal M}_k}} {{{\left( {\textit{var}\left( {{{\widehat \theta }_{\textit{im}}}} \right)} \right)}^2}} }},
\end{equation}
${{\cal M}_k} = {\cal M}/{{\cal A}_k}$, ${\cal M} = \left\{ {m|\forall m = 1,2, \cdots ,M} \right\}$, ${{\cal A}_k} = \textit{Unique}\left( {\left\{ {m_i^* = \arg \mathop {\max }\limits_m \textit{var}({{\widehat \theta }_{\textit{ij}}})\left| {\forall m \ne k} \right.} \right\}} \right)$, and $\textit{Unique}\left( {\cal T} \right)$ returns the same values as in set $\cal T$ but with no repetitions.

\section{Numerical Analysis}
As discussed in the introduction, the achievable rates in massive MIMO system with massive connectivity have been addressed in \cite{LiuYu2018a}. With the state evolution equations of AMP algorithm and large SNR approximation, the converged noise variance of the decoupled received pilot signal is determined as
\begin{equation}\label{equ26}
\tau _\infty ^2 = \frac{{{\sigma_0 ^2}}}{{1 - K/L}}
\end{equation}
where $K$ is the number of active users. Then, based on $\tau _\infty ^2$ and the MMSE channel estimate, uplink achievable rates with different detectors at the base station are identified. $\tau _\infty ^2$ obtained in \cite{LiuYu2018a} is in the same position as $\sigma _{\textit{eff}}^2$ in (\ref{equ5}). However, the use of (\ref{equ26}) requires the satisfaction of two conditions, namely large SNR and $K<L$, while our method to obtain $\sigma _{\textit{eff}}^2$ with the fixed point equation in (\ref{equ13}) does not need these conditions and hence is applicable more generally.

In this section, we compare the achievable uplink rates with both $\tau _\infty ^2$ and $\sigma _{\textit{eff}}^2$. Moreover, we can also obtain the noise variance of the decoupled received pilot signal numerically, and we refer to the achievable uplink rate with this method as the GAMP simulation result. Additionally, if the user activity is known a priori, we can assign orthogonal pilots to all the active users, and activity detection phase is not needed. It is obvious that the achievable uplink rate with prior knowledge on user activity provides an upper bound for the achievable rate without such information on user activity. As we discussed before, when $M$ tends to infinity, the error rates in activity detection tend to zero, which is equivalent to the case in which the prior information for user activity is available. Therefore, as $M$ grows without bound, the achievable uplink rate with $\sigma _{\textit{eff}}^2$ should overlap with the achievable rate with known user activities. Indeed, numerical results in this section verify this conclusion.

We assume that there are 1000 users uniformly distributed in a circular region with radius 500m. The path-loss decay exponent is $\alpha=2.5$. There are 1000 symbols in a channel coherence interval, and the bandwidth of the system is 1 MHz. The sum rate is averaged over $10^5$ realizations of this setting.

\begin{figure}[htbp]
	\center
	\includegraphics[width=3.2in]{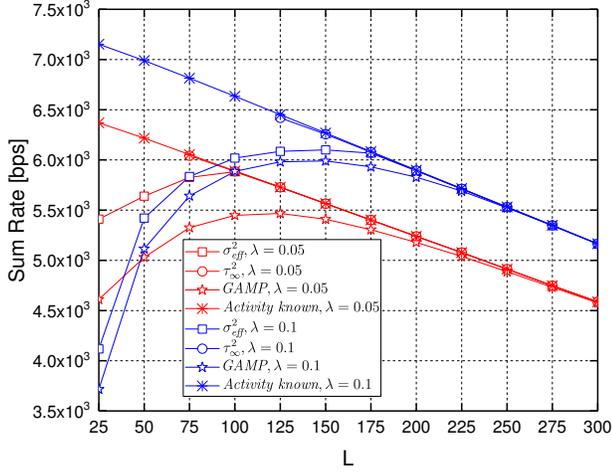}\\
	\caption{Achievable uplink rate versus the number of pilots 
}\label{fig1}
\end{figure}

Fig. \ref{fig1} plots the curves for the achievable uplink rate versus the number of pilots used in the activity detection and channel estimation phase, for different activity probabilities with 100 APs uniformly distributed in the circular region and SNR $=30$ dB. Since we can obtain $\tau _\infty ^2$ only when $L>K$, the sum rates based on $\tau _\infty ^2$ are plotted after $L=75$ when $\lambda=0.05$ and $L=125$ when $\lambda=0.1$. Since the number of symbols in a channel coherence interval is fixed, fewer symbols will be used in the uplink data transmission phase if more symbols are used for activity detection and channel estimation. Therefore, the sum rate of the system decreases with increasing $L$ when user activities are known at the CPU. On the other hand, a tradeoff exists when user activity needs to be detected. Specifically, when $L$ is small, error probability in activity detection is relatively high, leading to low achievable sum rates. Hence, a reduction in the sum rate is experienced due to uncertainty in user activity detection. Since the error probability in activity detection diminishes as $L$ increases, sum rates initially increase. However, as $L$ exceeds a threshold, smaller duration of time being available for uplink data transmission starts being the dominant factor and sum rates start decreasing. Therefore, there exists a tradeoff between sum rate and the number of pilots in the design of practical cell-free massive MIMO systems with massive connectivity and uncertainty in user activity. Comparing the curves for $\lambda=0.05$ and $\lambda=0.1$, we observe that the performance gap due to user activity uncertainty is smaller in the former case (due to smaller error rates in the presence of lower activity probabilities), while larger sum rates are achieved in the latter case (with $\lambda =0.1$) due to generally higher number of active users. We also notice that when $L$ is sufficiently large, the achievable sum rates based on $\sigma _{\textit{eff}}^2$ and $\tau _\infty ^2$ and also the sum rate obtained via GAMP simulations all start overlapping with that achieved when user activities are perfectly known a priori. Hence, in all cases, performance upper bound is approached as $L$ grows.

\begin{figure}[htbp]
	\center
	\includegraphics[width=3.2in]{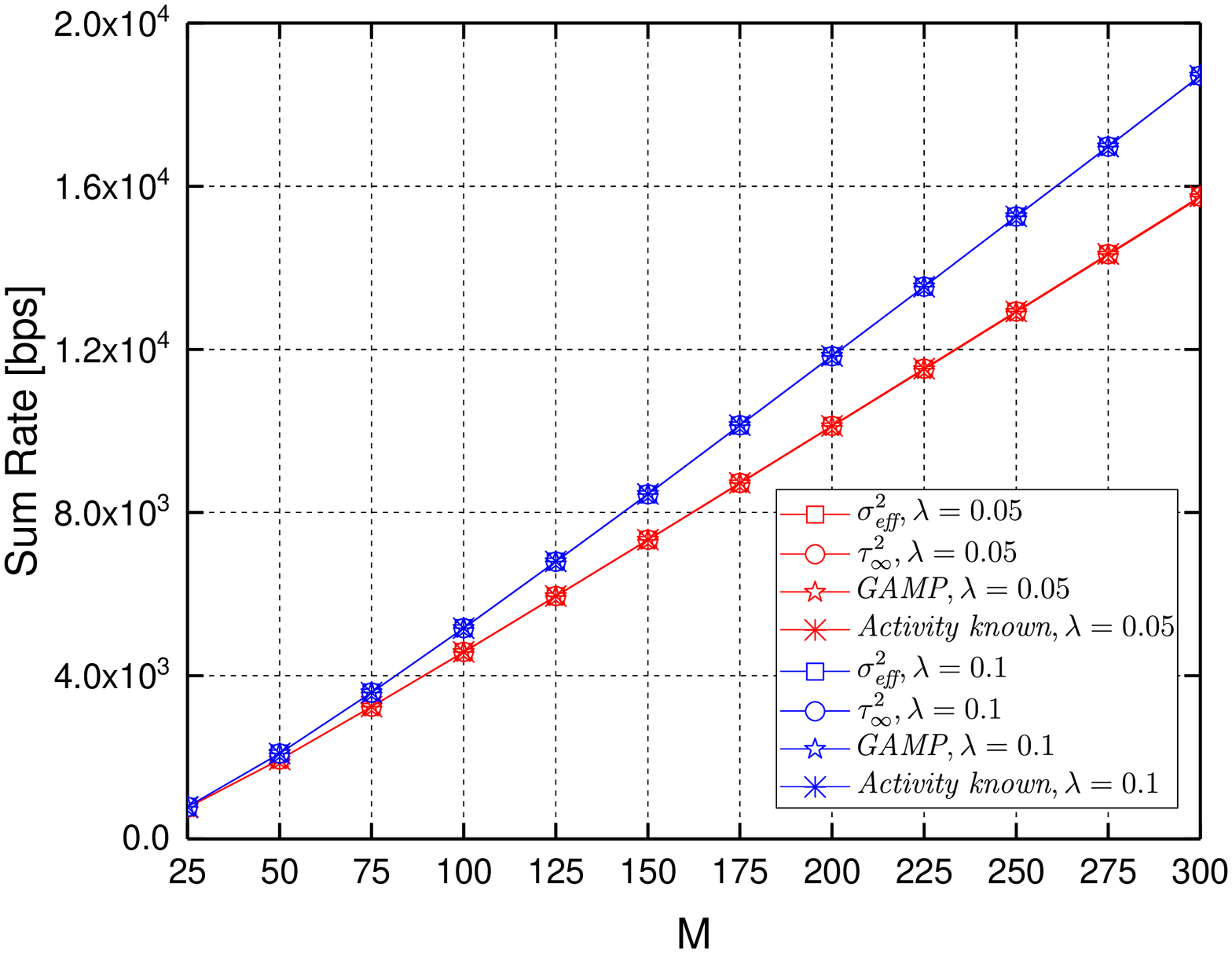}\\
	\caption{Achievable uplink rate versus the number of APs.}\label{fig2}
\end{figure}

Fig. \ref{fig2} plots the curves for achievable uplink rates versus the number of APs, for different activity probabilities with $L = 300$ and SNR $=30$ dB. These curves demonstrate that the sum rates based on $\sigma _{\textit{eff}}^2$ and $\tau _\infty ^2$ overlap with that achieved with prior information on user activities (labeled as ``activity known") and also with GAMP simulation results. This shows that the influence of user activity uncertainty on the achievable sum rate can be safely disregarded under assumption of relatively large $M$, $L$ and $N$ (e.g., $L = 300$ in this figure), i.e., having no errors in activity detection is a valid assumption. We also see that the achievable sum rates linearly increase as the number of APs grows, and the slope becomes higher as the activity probability increases. Furthermore, the achievable sum rates, similarly as before, diminish when $\lambda$ is decreased from 0.1 to 0.05.

\begin{figure}[htbp]
	\center
	\includegraphics[width=3.2in]{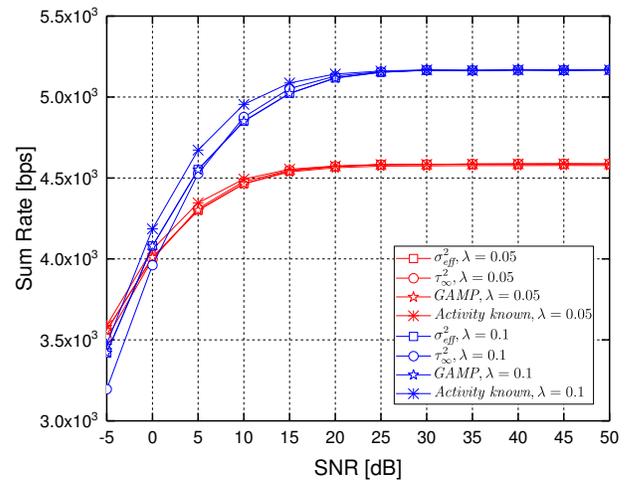}\\
	\caption{Achievable uplink rate versus SNR at APs.}\label{fig3}
\end{figure}

Fig. \ref{fig3} plots the curves for the achievable uplink rates versus SNR at APs for different activity probabilities with $L=300$ and $M=100$. When SNR is small, the error probability in activity detection is relatively large, which leads to the gaps between achievable sum rates with prior information on user activities and those achieved with $\sigma _{\textit{eff}}^2$, $\tau _\infty ^2$ and GAMP algorithm. These performance gaps can be regarded as the cost of uncertainty in the detection of user activity. We also notice that the achievable sum rates obtained with $\sigma _{\textit{eff}}^2$ and GAMP algorithm overlap, while a smaller sum rate is achieved with $\tau _\infty ^2$ when SNR is low. This verifies that $\sigma _{\textit{eff}}^2$ predicts the converged noise variance of GAMP algorithm better than $\tau _\infty ^2$ at low SNR levels, which is expected due to the fact that $\tau _\infty ^2$ is derived under the large SNR assumption. We further observe in the figure that the achievable sum rates increase and the cost of uncertainty in the detection of user activity diminishes as SNR grows. Indeed, when SNR is sufficiently large, the sum rates based on $\sigma _{\textit{eff}}^2$, $\tau _\infty ^2$ and GAMP algorithm overlap with that achieved with prior information on user activities. Similarly as before, when $\lambda$ is decreased from 0.1 to 0.05, both the achievable sum rates and the performance gap due to user activity uncertainty diminish.

\section{Conclusion}
Performance of cell-free massive MIMO systems with massive connectivity is analyzed in this paper. In the joint activity detection and channel estimation phase, MMSE estimates of the effective channel coefficients from all users to the entire set of APs are obtained via the GAMP algorithm. Then, the effective noise variance is determined with the state evolution equations of the GAMP algorithm. Following this characterization, the variances of both the estimated channel and channel estimation error are identified. Finally, the achievable uplink rates in the data transmission phase with ZF detector deployed at the CPU are analyzed and the impact of the number of pilots, the number of APs, SNR, and the user activity detection and channel estimation results on the performance is determined.

\end{spacing}

\bibliographystyle{IEEEtran}
\bibliography{rate}

\end{document}